\documentclass[aip,reprint,amsmath,amssymb]{revtex4-1}

\usepackage{graphicx}
\usepackage{dcolumn}
\usepackage{float}
\usepackage{bm}
\usepackage{amssymb}

\begin{document}

\title{Polymer translocation into cavities: Effects of confinement geometry,
crowding and bending rigidity on the free energy}

\author{James M. Polson and David R. Heckbert}
\affiliation{ Department of Physics, University of Prince Edward Island, 
550 University Ave., Charlottetown, Prince Edward Island, C1A 4P3, Canada }

\date{\today}

\begin{abstract}
Monte Carlo simulations are used to study the translocation of a polymer into a cavity.
Modeling the polymer as a hard-sphere chain with a length up to $N$=601 monomers,
we use a multiple-histogram method to measure the variation of the conformational free 
energy of the polymer with respect to the number of translocated monomers. 
The resulting free-energy functions are then used to obtain the confinement free energy for the 
translocated portion of the polymer. We characterize the confinement free energy for
a flexible polymer in cavities with constant cross-sectional area $A$ for various 
cavity shapes (cylindrical, rectangular and triangular) as well as for tapered cavities 
with pyramidal and conical shape. The scaling of the free energy with cavity volume and 
translocated polymer subchain length is generally consistent with predictions from simple 
scaling arguments, with small deviations in the scaling exponents likely due to finite-size
effects. The confinement free energy depends strongly on cavity shape anisometry and is
a minimum for an isometric cavity shape with a length/width ratio of unity.
Entropic depletion at the edges or vertices of the confining cavity
are evident in the results for constant-$A$ and pyramidal cavities. For translocation
into infinitely long cones, the scaling of the free energy with taper angle is consistent
with a theoretical prediction employing the blob model. We also examine the effects 
of polymer bending rigidity on the translocation free energy for cylindrical cavities. 
For isometric cavities, the observed scaling behaviour is in partial agreement with 
theoretical predictions, with discrepancies arising from finite-size effects that 
prevent the emergence of well-defined scaling regimes. In addition, translocation into 
highly anisometric cylindrical cavities leads to a multi-stage folding process for stiff 
polymers. Finally, we examine the effects of crowding agents inside the cavity. We find 
that the confinement free energy increases with crowder density. At constant
packing fraction the magnitude of this effect lessens with increasing crowder 
size for crowder/monomer size ratio $\geq 1$.
\end{abstract}

\maketitle

\section{Introduction}
\label{sec:intro}

The equilibrium conformational behaviour of polymers confined to small spaces has been the subject of 
much theoretical interest for decades.\cite{daoud1977statistics,deGennes_book} The basic
concept is straightforward: if one or more confinement dimensions is smaller than the 
mean size of the polymer, the number of accessible conformations is significantly reduced. 
This results a reduction in the conformational entropy and to an increase in the free energy of 
the polymer relative to its unconfined state. In spite of this apparent simplicity, theoretical
and computational studies have revealed a wide variety of scaling regimes for polymers 
confined to channels\cite{dai2016polymer} and cavities.\cite{gao2014free,sakaue2018compressing,%
sakaue2007semiflexible} Although the precise scaling behaviour is dependent on just a few system 
properties such as the confinement dimensions and the polymer contour length and bending 
rigidity, new regimes continue to be discovered.\cite{hayase2017compressive}
These theoretical studies have been complemented by progress in the experiment realm, 
where recent advances in nanofabrication techniques have enabled the study of DNA 
(deoxyribonucleic acid) confined to narrow channels\cite{reisner2012dna} 
and cavities.\cite{langecker2011electrophoretic,liu2015entropic,sampath2016dna,%
cadinu2017single,cadinu2018double,zhang2018single} In addition, some experimental studies have 
examined more complex confinement behaviour for cases where individual DNA molecules are distributed 
among many cavities connected by nanopores\cite{nykypanchuk2002brownian} or by narrow slits between
confining surfaces.\cite{klotz2015measuring,klotz2015correlated} The insights provided by
these experimental and theoretical studies are expected to benefit the development of nanofluidic 
technologies for the manipulation and analysis of DNA and other biopolymers.

Confinement is a relevant factor in many examples of polymer translocation through nanopores.%
\cite{Muthukumar_book}  For example, many biological phenomena such as viral DNA packaging or 
ejection, transport of messenger RNA (ribonucleic acid) across the nuclear pore complex and horizontal 
gene transfer between bacteria involve translocation into or out of a confined or otherwise crowded 
environment.\cite{Alberts_book,Lodish_book} In addition, some recently developed experimental 
techniques for studying translocation use 
devices that incorporate confinement of DNA in cavities. For example, Liu {\it et al.} designed a
device with an ``entropic cage'' placed near a solid-state nanopore to trap a translocated DNA 
molecule.  Upon chemical modification inside the cage the same molecule can be driven back through the 
pore and a comparison of the ionic current traces for translocation enables characterization of
the altered DNA.\cite{liu2015entropic} Langecker {\it et al}. measured the mobility of a DNA molecule
using time-of flight measurements with a stacked-nanopore device in which the molecule enters a 
pyramidal cavity through one pore and exits through a second pore.\cite{langecker2011electrophoretic}
Another recent study examined translocation into conical enclosures.\cite{bell2017asymmetric}
Optimizing the functionality of such devices would benefit from an understanding of the 
effects of cavity shape and size on the translocation process.

Numerous theoretical and computer simulation studies have examined polymer translocation
into or out of confined spaces of various geometries, including spherical or ellipsoidal 
cavities\cite{muthukumar2001translocation,muthukumar2003polymer,kong2004polymer,ali2004dynamics,%
ali2005coarse,cacciuto2006confinement,ali2006polymer,forrey2006langevin,ali2008ejection,%
sakaue2009dynamics,matsuyama2009packaging,ali2011influence,yang2012adsorption,%
rasmussen2012translocation,ghosal2012capstan,zhang2012dynamics,al2013effect,zhang2013dynamics,%
polson2013simulation,polson2013polymer,mahalik2013langevin,linna2014dynamics,zhang2014polymer,%
cao2014dynamics,piili2015polymer,polson2015polymer,linna2017rigidity,piili2017uniform,%
sun2018theoretical,sun2019trapped}, cylindrical cavities,\cite{sean2017highly} or laterally 
unbounded spaces between flat walls.\cite{luo2009polymer,luo2010polymer,luo2011chain,%
sheng2012ejection,yang2016semiflexible} Many of these studies have emphasized on the role 
of the confinement free energy in driving polymer translocation out of the enclosure or 
in countering other applied forces that drive polymers into such
spaces.\cite{muthukumar2001translocation,muthukumar2003polymer,kong2004polymer,ali2004dynamics,%
cacciuto2006confinement,ali2006polymer,matsuyama2009packaging,yang2012adsorption,%
rasmussen2012translocation,zhang2012dynamics,yang2012adsorption,zhang2013dynamics,%
rasmussen2012translocation,polson2013polymer,sun2018theoretical} One approach to
interpreting the observed dynamics is using the Fokker-Planck (FP) formalism with the 
translocation free-energy functions.\cite{Muthukumar_book} Although recent theories
of polymer translocation have emphasized the importance of out-of-equilibrium effects
on the translocation dynamics,\cite{panja2013through,palyulin2014polymer} it has
been noted by Katkar and Muthukumar\cite{katkar2018role} that numerous experimental 
studies have reported results consistent with quasistatic translocation, a condition
required for the valid application of the FP formalism. Consequently, the characterization
of the translocation free-energy functions is of value. 

Of the simulation studies that have examined translocation into or out of cavities, most have
focused on spherical cavities while only a few have considered the effects of cavity shape 
anisometry.\cite{ali2006polymer,zhang2013dynamics,zhang2014polymer,polson2015polymer,sean2017highly}
In addition, studies in which direct calculation of the confinement free energy using Monte Carlo 
methods have been carried out typically address only the simple case of spherical 
cavities.\cite{cacciuto2006self} 
Given the variety of confinement cavity shapes used in the recent DNA translocation experiments 
described above, it is clear that characterization of the free energy with respect to cavity shape 
would be useful. In a recent study, we made some progress toward this goal. Using
a multiple-histogram MC method we measured the variation in the translocation free-energy 
function for the case of ellipsoidal cavities and observed a significant effect on the confinement
free energy by varying the cavity anisometry.\cite{polson2015polymer} Generally, for a given
cavity volume, we found that the free energy is lowest for spherical cavities and increases as the 
cavity shape becomes more oblate or prolate. The purpose of the present study is to extend that work.
We consider here cavities of a variety of shapes, including cylindrical, rectangular and
triangular, as well as those with tapered geometries such as cones and pyramids. We also
consider the effects of varying the polymer bending rigidity as well as the presence of crowding 
agents inside the cavity.  The scaling properties of the free-energy functions are compared
with predictions using simple models and recent theoretical studies. Generally, the results 
are semi-quantitatively consistent with the predictions, with small discrepancies between 
measured and predicted scaling exponents likely arising from finite-size effects.

The remainder of this article is organized as follows. Section~\ref{sec:model} presents a 
brief description of the model employed in the simulations, following which 
Section~\ref{sec:methods} gives an outline of the methodology employed and
other relevant details of the simulations. Section~\ref{sec:results} presents the simulation 
results for the various systems we have examined.  Finally, Section~\ref{sec:conclusions} 
summarizes the main conclusions of this work.

\section{Model}
\label{sec:model}

We employ a minimal model to describe a polymer translocating through a nanopore in a flat
barrier from a semi-infinite space into a cavity.
The polymer is modeled as a chain of hard spheres, each with diameter $\sigma$.  
The pair potential for non-bonded monomers is thus $u_{\rm{nb}}(r)=\infty$ for $r\leq\sigma$ 
and $u_{\rm{nb}}(r)=0$ for $r>\sigma$, where $r$ is the distance between the centers of the 
monomers. Pairs of bonded monomers interact with a potential $u_{\rm{b}}(r)= 0$ if 
$0.9\sigma<r<1.1\sigma$ and $u_{\rm{b}}(r)= \infty$, otherwise. In the case of semiflexible 
polymers, the stiffness of the chain is modeled using a bending potential associated with 
each consecutive triplet of monomers. The potential has the form, 
$u_{\rm bend}(\theta) = \kappa (1-\cos\theta)$.  The angle $\theta$ is defined at monomer
$i$ such that $\cos\theta_i \equiv \hat{u}_{i}\cdot\hat{u}_{i+1}$, where $\hat{u}_i$ is a 
normalized bond vector pointing from monomer $i-1$ to monomer $i$. The bending constant 
$\kappa$ determines the stiffness of the polymer and is related to the persistence length 
$P$ by $\kappa/k_{\rm B}T = P/\langle l_{\rm bond}\rangle\approx P/\sigma$, as the mean 
bond length is $\langle l_{\rm bond} \rangle \approx \sigma$.

\begin{figure}[!ht]
\begin{center}
\vspace*{0.2in}
\includegraphics[width=0.35\textwidth]{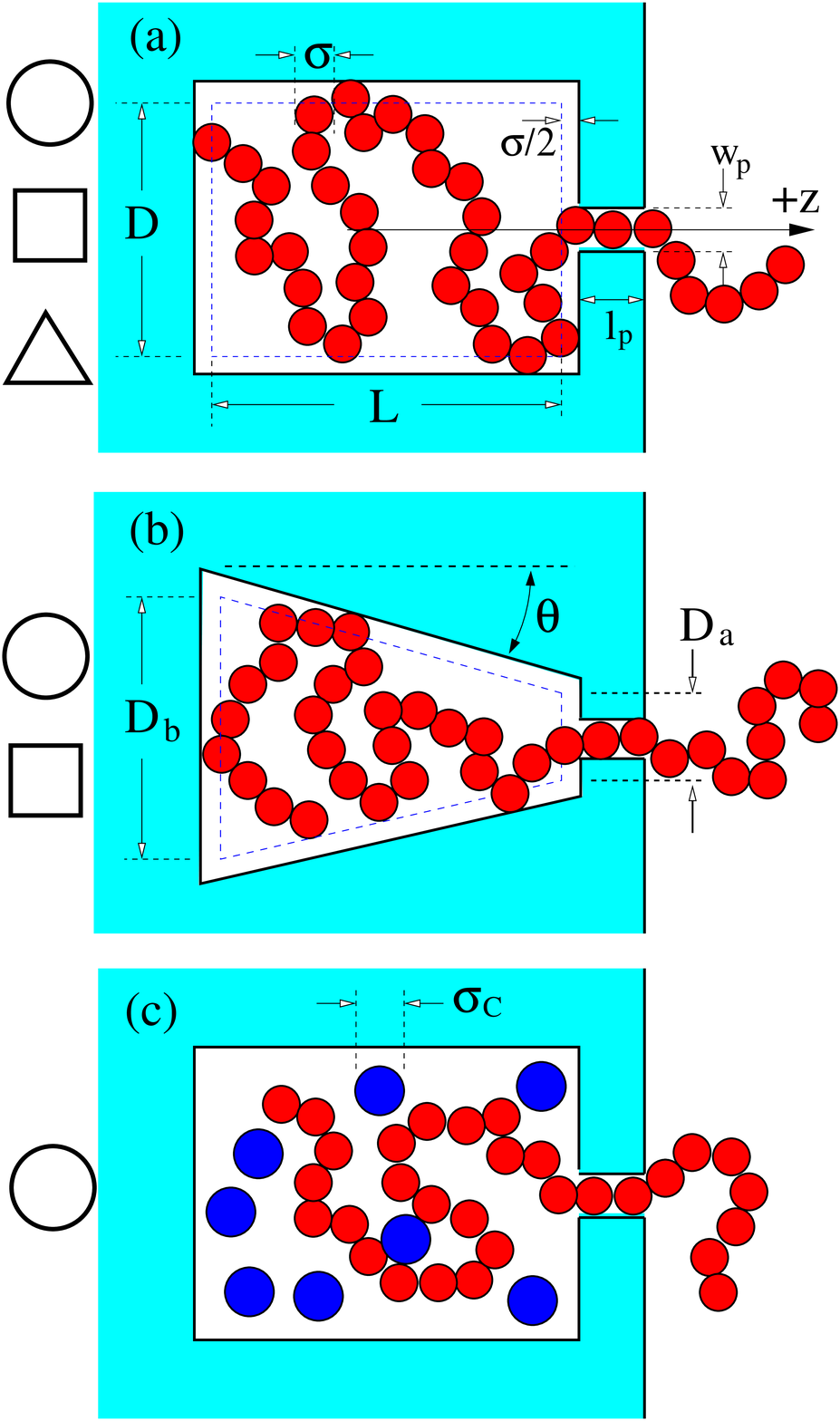}
\end{center}
\caption{Illustration of the system studied in this work. (a) The polymer translocates
through a cylindrical pore of length $l_{\rm p}$ and diameter $w_{\rm p}$ from a cavity
of dimensions $L$ and $D$, defined by the subspace accessible to the centers of the monomers.
The lateral dimension is defined $D=\sqrt{A}$, where $A$ is the constant cross-sectional
area of the subspace. (b) As in (a) except the cavity has a varying cross-sectional area.
The lateral dimensions of the subspace at the truncated apex and base of the cavity are 
$D_{\rm a}$ and $D_{\rm b}$, respectively. (c) As in (a), except crowding agents of size
$\sigma_{\rm c}$ partially occupy the cavity space.  The shapes on the left side of each
picture indicate the cavity cross-section shapes examined in each case.}
\label{fig:illust}
\end{figure}

We consider confinement cavities of two main types. In the first case, we consider
cavities with constant cross-sectional area that have circular, square and
(equilateral) triangular cross sections.  In the second case we examine
tapered cavities with variable cross-sectional area with circular and square cross 
sections, which correspond to conical and pyramidal shaped spaces. For these cavities,
the cone or pyramid is truncated at the apex. The walls of each cavity are ``hard''
such that the monomer-wall interaction energy is $u_{\rm w}(r) = 0$ if monomers do not 
overlap with the wall and $u_{\rm w}(r) = \infty$ if there is overlap. The effective 
channel width is defined to be $D=\sqrt{A}$, where $A$ is the cross-sectional 
area for the subspace in the channel that is accessible to the centers of the monomers. 
Likewise, the cavity length $L$ is measures the span of the same subspace.
The aspect ratio of the cavity is defined $r\equiv L/D$.  The conical 
and pyramidal cavities are characterized by two effective widths, $D_{\rm a}$ and 
$D_{\rm b}$ ($>D_{\rm a}$) at the (truncated) apex and the base, respectively. 
A single nanopore of length $l_{\rm p}$ and with $w_{\rm p}$ is located on one end
of the cavity. In most of the calculations we use $l_{\rm p}=1.4\sigma$ and $w_{\rm p}=1.4\sigma$.
In some simulations we include crowding agents, which are modeled as hard spheres
with a diameter of $\sigma_{\rm c}$ and are confined to the cavities. 
The various model systems are illustrated in Fig.~\ref{fig:illust}.

Note that unlike Ref.~\onlinecite{polson2015polymer} we do not include forces required to actually 
drive the polymer into the cavity (such as electric forces in the pore or attraction to the cavity 
surface).  Consequently, the free energy is greater inside the cavity than it is outside. Thus, 
the polymer is spontaneously driven outward from the cavity for the model systems used here. Inclusion 
of forces that offset the effect of this free energy gradient to drive translocation inward
is straightforward but beyond the scope of this study.

\section{Methods}
\label{sec:methods}

Monte Carlo simulations employing the Metropolis algorithm and the self-consistent histogram 
(SCH) method\cite{frenkel2002understanding} were used to calculate the free-energy functions
for the polymer-nanopore model described in Section~\ref{sec:model}. The SCH method provides 
an efficient means to calculate the equilibrium probability distribution ${\cal P}(m)$, and 
thus its corresponding free-energy function, $F(m) = -k_{\rm B}T\ln {\cal P}(m)$. 
Here, $m$ is defined as the number of bonds that have crossed the mid-point of the nanopore. 
Typically, one bond spans this point for any given configuration, and this bond contributes to $m$
the fraction that lies on the cavity side of the point.  Note that $m$ is a continuous
variable in the range $m\in[0,N-1]$, and $m$$-$1 is essentially the number of monomers inside
the cavity. We have 
previously used this procedure to measure free-energy functions in other simulation studies 
of polymer translocation\cite{polson2013simulation,polson2013polymer,polson2014evaluating}
as well in studies of polymer segregation under cylindrical 
confinement\cite{polson2014polymer,polson2018segregation} and polymer folding in long
nanochannels.\cite{polson2017free,polson2018free}

To implement the SCH method, we carry out many independent simulations, each of which employs a
unique ``window potential'' of a chosen functional form.  The form of this potential is given by:
\begin{eqnarray}
{W_i(m)}=\begin{cases} \infty, \hspace{8mm} m<m_i^{\rm min} \cr 0,
\hspace{1cm} m_i^{\rm min}<m<m_i^{\rm max} \cr \infty, \hspace{8mm} m>m_i^{\rm max} \cr
\end{cases}
\label{eq:winPot}
\end{eqnarray}
where $m_i^{\rm min}$ and $m_i^{\rm max}$ are the limits that define the range of $m$
for the $i$-th window.  Within each ``window'' of $m$, a probability distribution $p_i(m)$ is
calculated in the simulation. The window potential width,
$\Delta m \equiv m_i^{\rm max} - m_i^{\rm min}$, is chosen to be sufficiently small
that the variation in $F$ does not exceed a few $k_{\rm B}T$. Adjacent windows overlap,
and the SCH algorithm uses the $p_i(m)$ histograms to reconstruct the unbiased distribution,
${\cal P}(m)$. The details of the histogram reconstruction algorithm are given in 
Refs.~\onlinecite{frenkel2002understanding} and \onlinecite{polson2013simulation}.

Polymer configurations were generated carrying out single-monomer moves using a combination of 
translational displacements and crankshaft rotations.  In addition, reptation moves were also 
employed.  The trial moves were accepted with a probability 
$p_{\rm acc}=\min(1,e^{-\Delta E/k_{\rm B}T})$, where $\Delta E$ is the energy difference 
between the trial and current states.  Prior to data sampling, the system was equilibrated.  
As an illustration, for a $N=601$ polymer chain, the system was equilibrated for typically 
$\sim 10^7$ MC cycles, following which a production run of $\sim 10^8$ MC cycles was carried 
out.  On average, during each MC cycle one reptation move and one single-monomer displacement 
or crankshaft rotation for each monomer is attempted once.

The windows are chosen to overlap with half of the adjacent window, such that $m^{\rm max}_{i} =
m^{\rm min}_{i+2}$.  The window width was typically $\Delta m = 4\sigma$. Thus, a calculation for 
$N=601$, where the translocation coordinate spans a range of $m\in[0,600]$, required separate 
simulations for 299 different window potentials.  For each simulation, individual probability 
histograms were constructed using the binning technique with 20 bins per histogram.

In the results presented below, distance is measured in units of $\sigma$
and energy in units of $k_{\rm B}T$. 

\section{Results}
\label{sec:results}

\subsection{Translocation of fully flexible polymers into isometric cavities}
\label{subsec:scalconf}

\begin{figure}[!ht]
\begin{center}
\vspace*{0.2in}
\includegraphics[width=0.45\textwidth]{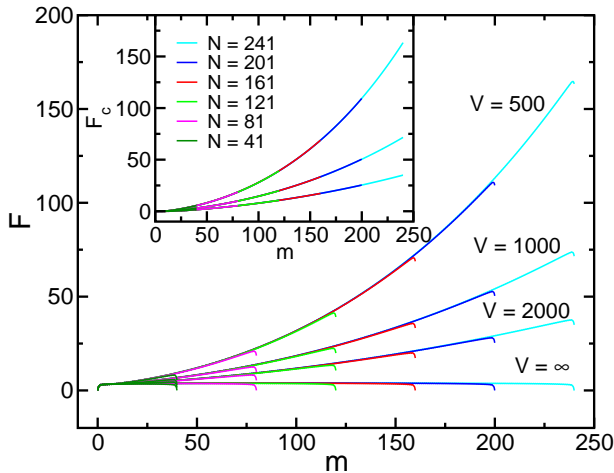}
\end{center}
\caption{Free energy functions for several polymer lengths ($N$=41, 81, 121, 161, 201 and 241) 
and a cylindrical cavity of aspect ratio $r=1$. Results for three different cavity volumes
are shown. The inset shows $F_{\rm c}$ vs $m$, where $F_{\rm c}(m) \equiv F(m;V)-F_0(m)$ and 
$F_0(m)\equiv F(m;V=\infty)$.}
\label{fig:F_length_vol}
\end{figure}

We consider first the scaling properties of the free energy of flexible polymers translocating 
into isometric cylindrical cavities, i.e. cylindrical cavities with an aspect ratio of 
$r\equiv L/D=1$. Figure~\ref{fig:F_length_vol} shows the variation of $F$ with translocation 
coordinate $m$ for various cavity volumes and polymer lengths. The free-energy curves are 
vertically shifted so that $F=0$ at $m=0$.  Note that the limiting case of $V$=$\infty$ 
corresponds to translocation between two semi-infinite subspaces through an infinitely large 
flat wall. In this case, $F$ is nearly constant with respect to $m$ with only slight decreases near 
the limiting values of $m=0$ and $m=N-1$.  The shape of this profile is well understood and in 
the case of infinite polymer length is given by $F(m) = (1-\gamma)\ln[(N-m)m]$, where 
$\lambda\approx 0.69$ is a critical exponent in 3D.\cite{Muthukumar_book} For cavities of finite 
volume, the free energy increases monotonically with $m$ (except near $m=N-1$). The rate 
of this increase of $F$ increases as the volume $V$ of the cavity space decreases. This 
follows from the fact that increasing confinement reduces the number of accessible 
conformations of the polymer and thus lowers the entropy. The free-energy functions each 
have positive curvature over most of their range. This results from the fact that as 
translocation proceeds, the fraction of the cavity space occupied by monomers increases. 
The reduction in available cavity space means that the loss in conformational
entropy upon transfer of each monomer from outside to the inside of the cavity also
increases.

Another notable feature in Fig~\ref{fig:F_length_vol} is the overlap of the curves for 
polymers of different contour lengths entering a cavity of a given volume. This overlap
arises from the fact that the confinement free energy of the translocated subchain of length 
$m$ dominates the total free energy, and the confinement free energy of this portion of
the polymer is independent of polymer contour length. We define the confinement free energy 
as the difference $F_{\rm c}(m) \equiv F(m)-F_0(m)$, where $F_0(m)\equiv F(m;V=\infty)$
is the free energy for polymer translocation through a flat barrier.
To clarify the meaning of $F_{\rm c}(m)$, consider the commonly used approximation 
that the free energy of a partially translocated polymer is the sum of contributions from 
two subchains, one of length $m$ on the {\it trans} side of the pore, and the other of length 
$N-m$ on the {\it cis} side, each of which is effectively tethered to the pore-containing 
wall.\cite{Muthukumar_book} Thus, $F(m) = F^{\rm (cw)}_{\rm trans}(m)+F^{\rm (cw)}_{\rm cis}(N-m)$ 
and $F_0(m) = F^{\rm (w)}_{\rm trans}(m)+F^{\rm (w)}_{\rm cis}(N-m)$,
where ``w'' denotes tethering to a wall and ``c'' denotes the presence of cavity
confinement. For the systems considered in this work illustrated in Fig.~\ref{fig:illust}
the confinement only affects the {\it trans} subchain. It follows that
$F^{\rm (cw)}_{\rm cis}(N-m) = F^{\rm (w)}_{\rm cis}(N-m)$, and so
$F_{\rm c}(m) = F^{\rm (cw)}_{\rm trans}(m) - F^{\rm (w)}_{\rm trans}(m)$.
Thus, $F_{\rm c}(m)$ can be interpreted as the additional free energy of a polymer 
tethered to a hard wall arising from a reduction in the conformational entropy due to
the cavity confinement. The approximation employed neglects subtle effects from the
pore that lead to oscillations in the free energy, as described in detail in 
Ref.~\onlinecite{polson2013simulation}. However, it can be shown that subtraction of
$F_0(m)$ from the free-energy function eliminates this feature in $F_{\rm c}(m)$.
The confinement free-energy functions for the data of Fig.~\ref{fig:F_length_vol} are
shown in the inset of the figure. Note that the small deviations from perfect 
overlap for $F(m)$ are now gone and the $F_{\rm c}(m)$ overlap perfectly for all
$N$ at each cavity volume, as expected. (The deviations from perfect overlap for the
$F(m)$ curves are due to the non-extensive part of the free energy of a wall-tethered
polymer, which also gives rise to the slight curvature in the free-energy curves in
the absence of confinement ($V$=$\infty$).)

\begin{figure}[!ht]
\begin{center}
\vspace*{0.2in}
\includegraphics[width=0.45\textwidth]{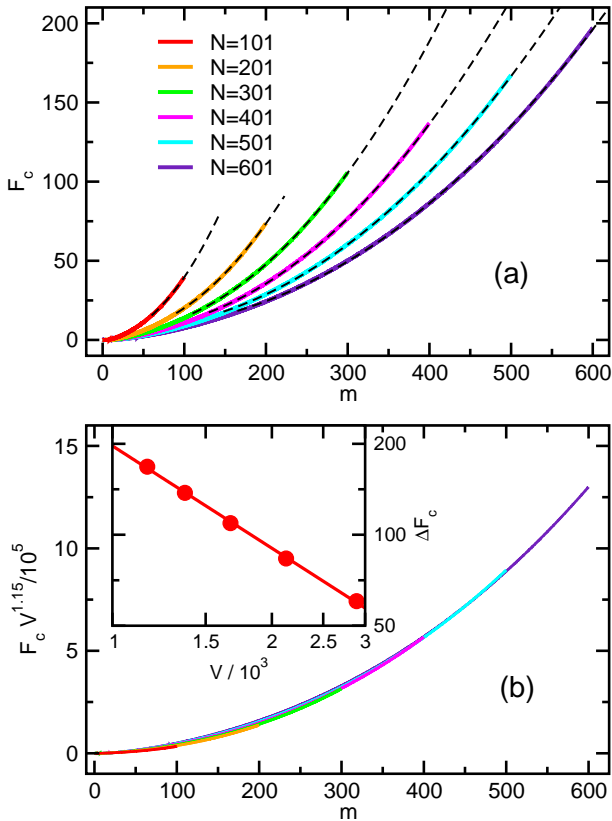}
\end{center}
\caption{(a) Free energy difference $F_{\rm c} = $ vs $m$ for polymers of different lengths.  
In each case, the confinement cavity is a cylinder with an aspect ratio of $D/L$=1 and a volume 
determined by the condition that the packing fraction for $m=N-1$ is $\phi_{\rm c}=0.15$. The 
dashed lines show fits to the curves for a fitting function of the form 
$F_{\rm c} = c_0+c_1(m-1)^\beta$. The minimum $m$ of the fitting curves is the minimum of 
the range over which the $F_{\rm c}$ was fit. The fitting exponents are $\beta$=2.01, 2.05, 
2.08, 2.10, 2.11 and 2.13 for $N$=101, 201, 301, 401, 501 and 601, respectively. 
(b) Scaled free-energy functions using the data from (a), where the scaling $F_{\rm c} V^{1.15}$ 
leads to data collapse, as explained in the text. The inset shows the variation of 
$\Delta F_{\rm c} (\equiv F_{\rm c}(m=N))$ vs cavity volume for a $N$=401 polymer entering a 
cavity with $D/L=1$. The solid line shows a fit to a power law $\Delta F_{\rm c}\sim V^\alpha$, 
where the fit yielded a scaling exponent of $\alpha$=$1.12\pm 0.02$.  }
\label{fig:delF_fit}
\end{figure}

Let us now examine the scaling properties of $F_{\rm c}(m)$.  Figure~\ref{fig:delF_fit} 
shows the confinement free energy for polymers with lengths ranging from $N$=101 to 601. 
For convenience, each simulation was carried out for a cavity volume chosen so that 
the packing fraction in the cavity at $m=N-1$ was $\phi=0.15$; consequently, the cavity 
volume was proportional to the polymer length.  The variation of $F_{\rm c}$ with $m$ can 
be estimated using scaling arguments developed for polymer solutions in the semi-dilute 
regime. Here, the confined section of the polymer can be viewed as a collection of blobs, 
each with a size of $\xi\sim \phi^{\nu/(1-3\nu)}$, where $\phi\sim m/V$ is the packing 
fraction and $\nu\approx 0.588$ is the Flory exponent.\cite{Rubinstein_book} Since the 
number of blobs is $n_{\rm b}\approx V/\xi^3$ and each blob contributes of order $kT$ to 
the confinement free energy, it follows that
\begin{eqnarray}
F_{\rm c}/kT \sim V^{-\alpha} (m-1)^{\beta},
\label{eq:FVm}
\end{eqnarray}
where $\alpha$=1.31 and $\beta$=2.31. Note that we have substituted $m\rightarrow m-1$ to 
account for the finite length of the pore, which holds approximately one monomer. Also 
note that the commonly employed approximation of $\nu=\frac{3}{5}$ leads to a slightly 
different scaling of 
$F_{\rm c}/kT \sim V^{-1.25} (m-1)^{2.25}$. Previous work has shown that the semi-dilute 
regime scaling is accurate for packing fractions of $\phi<0.15$,\cite{cacciuto2006self}, 
which is the motivation here for choosing the cavity volume to be such that $\phi$=0.15 at 
full insertion.  As a consequence, the condition that $\phi<0.15$ is satisfied for all $m$. 
A lower limit on the range of validity for this prediction is the requirement that the number 
of blobs $n_{\rm b}\approx (m-1)\phi^{1/(3\nu-1)}$ satisfy $n_{\rm b}\gg 1$. For the polymer 
lengths considered here, it is not possible to find a range of $m$ that satisfies both conditions 
simultaneously. To analyze the data, we follow the approach taken in our previous 
study\cite{polson2015polymer} and use the more relaxed condition for low density of 
$n_{\rm b}\geq 3$. 

Figure~\ref{fig:delF_fit}(a) shows the results of fits to each of the confinement free-energy
functions using a fitting function of the form $F_{\rm c}=c_0+c_1(m-1)^\beta$, where $c_0$,
$c_1$ and $\alpha$ are fitting coefficients. The best fit curves are plotted on the graph
as dashed lines. The lower bounds of the plotted fitting curves mark the lower limit of the
range of the simulation data that were included in the fit, i.e. the point where $n_{\rm b}=3$.
The fitting exponents were measured to be $\beta$=2.01, 2.05, 2.08, 2.10, 2.11 and 2.13 for
$N$=101, 201, 301, 401, 501 and 601, respectively. These values are underestimates of the
predicted scaling exponent of $\beta=2.31$. This is clearly a finite-size effect, as suggested
by the fact that $\beta$ tends (slowly) toward the predicted value as the system size increases.
As noted above, a different cavity volume was used for each simulation. Equation~(\ref{eq:FVm}) 
predicts that the scaling $F_{\rm c}V^{1.31}$ should collapse these functions onto a universal 
curve if we scale using the volume $V$ employed in each of the simulations.  As is evident in 
Fig.~\ref{fig:delF_fit}(b), we find scaling with somewhat smaller exponent, i.e. 
$F_{\rm c}V^{1.15}$, produces the best collapse. As with the small discrepancy in the 
observed and predicted variation of $F_{\rm c}$ with respect to $m$, this difference is 
undoubtedly due to finite-size effects.  
To further investigate the volume dependence of $F_{\rm c}$, we examine the case of translocation 
of a $N$=401 polymer into a $r$=1 cavity of different volumes. We define the confinement
free energy at full insertion to $\Delta F_{\rm c}\equiv F_{\rm c}(m$=$N$$-$$1)$. The inset 
of Fig.~\ref{fig:delF_fit}(b) shows the variation of $\Delta F_{\rm c}$ with $V$ in cases 
where the cavity volume fraction at full insertion was $\phi_{\rm c}$=0.06, 0.08, 0.1, 0.12 
and 0.14. Fits to a power law yield a scaling of $\Delta F_{\rm c}\sim V^{-1.12\pm 0.02}$.
Again we find a discrepancy between the measured exponent and that predicted using scaling 
arguments.  

The discrepancies between the measured and predicted scaling exponents $\alpha$ and $\beta$ 
are somewhat smaller than those obtained in our previous study of translocation into spherical 
cavities (a special case of ellipsoidal cavities that were studied).\cite{polson2015polymer} 
This is likely due to the shorter polymer lengths considered in that study ($N \leq 140$) and 
further supports our claim that they are due to finite-size effects. 

It is worth noting here that the confinement free energy calculated in the simulations
is that for a polymer whose end monomer is effectively tethered to a point on the inner
surface of the confining cavity. This is due to the fact that when $m$=$N$-$1$, a single
monomer is still located in the nanopore. On the other hand, the theoretical model imposes
no such condition. In principle, the difference in confinement free energies for these
two cases will contribute to the discrepancy. In a recent study, we described a method
to calculate the free-energy cost of localizing an end monomer of a confined polymer.\cite{polson2019free}
Using the same model for flexible chains as that employed here we find that for polymer lengths
and packing fractions comparable to those used here that the end-monomer localization
free energy was 1--2~$k_{\rm B}T$. Thus, the effect is very small and unlikely to be
the principal cause of the discrepancy.

Why do finite-size effects lead to effective scaling exponents that are lower than 
the predicted values? Some insight is provided by the arguments presented by Sakaue in
Ref.~\onlinecite{sakaue2007semiflexible}. The scaling prediction derived above begins
with the assumption that $F_{\rm c}\sim n_{\rm blob} = V/\xi^3$, which implicitly assumes
that the monomer density is uniform throughout the enclosed space. However, Sakaue notes
that monomer depletion in a layer of width $\approx\xi$ near the surface of the cavity 
is expected. This gives rise to a surface correction term to the free energy, 
$\Delta F_{\rm surf}$, which is approximated as a surface integral 
$\Delta F_{\rm surf} = \int da (\Pi\times \xi)$, where the osmotic pressure is given by
$\Pi\approx k_{\rm B}T/\xi^3$. This is approximately $\Delta F_{\rm surf} = \Pi\xi A_{\rm s}$,
where $A_{\rm s}\sim D^2$ is the surface area of the cavity. Noting again that 
$\xi\sim \phi^{1/(1-3\nu)}$, it follows that 
\begin{eqnarray}
\Delta F_{\rm surf}/k_{\rm B}T \sim D^{-0.87} (m-1)^{1.54}.
\label{eq:FVm2}
\end{eqnarray}
(Using $\nu=\frac{3}{5}$ gives $\Delta F_{\rm surf}/k_{\rm B}T \sim D^{-0.83} (m-1)^{1.5}$.)
Comparing Eqs.~(\ref{eq:FVm}) and (\ref{eq:FVm2}), we see that the exponents for $D$ and
$m-1$ in the case of the surface term $\Delta F_{\rm surf}$ are each smaller than those for the
volume term (i.e. $0.87 < 1.31$ and $1.54 < 2.31$). For a sufficiently small cavity, the 
surface term could be an appreciable contribution to the total free energy. For the results 
shown in Fig.~\ref{fig:delF_fit}, this appears to be the case, and the smaller scaling
exponents of the surface term reduce the values of the measured effective exponents.
A thorough investigation of this effect requires additional calculations with much larger
cavity sizes where the volume of the depletion layer near the surface is a much smaller
fraction of the entire cavity volume. However, this also necessitates using polymers
at least an order of magnitude longer, which is currently not feasible.

\subsection{Effects of confinement shape}
\label{subsec:shape}

\begin{figure}[!ht]
\begin{center}
\vspace*{0.2in}
\includegraphics[width=0.45\textwidth]{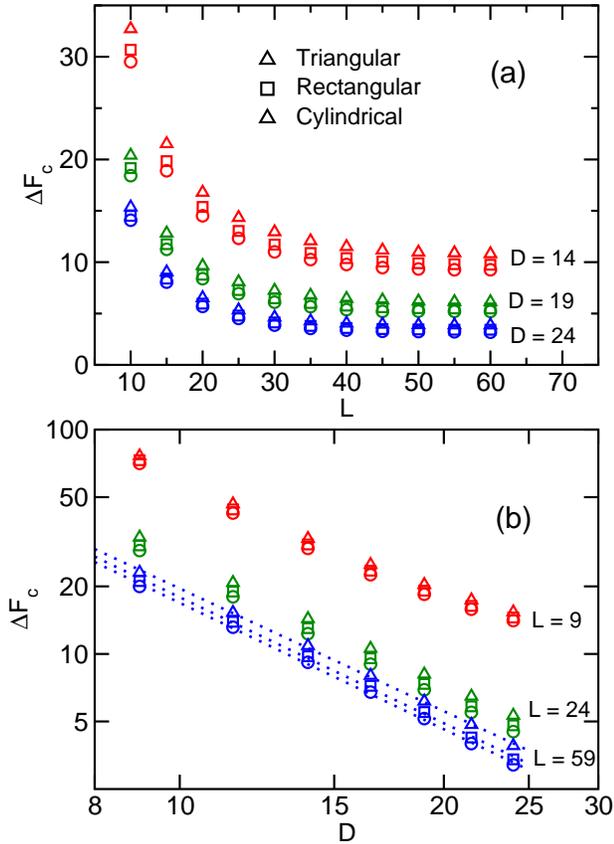}
\end{center}
\caption{(a) Confinement free energy $\Delta F_{\rm c}$ vs cavity length for various cavity 
geometries. Results are shown for a polymer of length $N$=201 for three different cavity widths.
(b) Confinement free energy $\Delta F_{\rm c}$ vs cavity width $D$ for various cavity
geometries. Results are shown for a polymer of length $N$=201 for cavity lengths of
$L$=9 (red symbols), $L$=24 (green symbols) and $L$=59 (blue symbols). The blue dashed
lines for $L$=59 are fits to a power law, $\Delta F_{\rm c}\sim D^{-\beta}$, which yield
exponents of $\beta=1.82\pm 0.03$ (triangular), $\beta=1.87\pm 0.03$ (rectangular),
and $\beta=1.87\pm 0.03$ (cylindrical).}
\label{fig:delF.L.D}
\end{figure}

We now consider how the free energy is affected by varying the both the geometry type and 
the shape anisometry of the cavity. In this section we examine translocation into cavities
with constant cross-sectional area for the cases of cylindrical, rectangular and triangular 
cross-sections as illustrated in Fig.~\ref{fig:illust}(a).  
Figure~\ref{fig:delF.L.D}(a) shows the confinement free energy $\Delta F_{\rm c}$ vs cavity 
length $L$ for three different cavity types, each with three different values of cavity 
width $D$. Results are shown for a polymer of length $N$=201.  Several trends are notable. 
First, for all cavity dimensions $D$ and $L$ the confinement free energy is greatest for 
triangular cavities and lowest for cylindrical cavities. This is due to the effect of 
entropic depletion, in which the monomer density is significantly reduced in sharp
corners of confined spaces.\cite{manneschi2013conformations,reinhart2013entropic,polson2018free} 
This effect tends to be especially strong in triangular cavities, which have the sharpest angles, 
and is absent for the case of cylindrical confinement. Such monomer depletion in these regions 
leads to an effective cross-sectional area that is less than the actual area accessible to 
the monomer centers. Since decreasing the area and therefore $D$ increases the confinement
and therefore the free energy, the trends with regard to cross-section shape follow
accordingly.

Another trend is that for sufficiently long channels $\Delta F_{\rm c}$ is invariant
with respect to $L$. This results simply from the fact that a polymer in a sufficiently 
long tube is insensitive to the presence of longitudinal confinement. Thus, decreasing
$L$ in this range does not reduce the number of accessible conformations and decrease
the entropy. However, as $L$ is further reduced the polymer is uniformly compressed
along the channel, leading to entropy loss and the observed increase in $\Delta F_{\rm c}$.
The onset of the effects of longitudinal confinement upon decreasing $L$ occurs at higher
values of $L$ for narrower channels, reflecting the fact that the extension length
is greater for smaller channels widths. In addition, $\Delta F_{\rm c}$ decreases with
increasing channel width at each fixed tube length. This is a consequence of the fact 
that narrower channels distort the polymer more relative to the unconfined state, leading
to a greater reduction in entropy.

Figure~\ref{fig:delF.L.D}(b) shows the variation of $\Delta F_{\rm c}$ with $D$ for 
the same three cavity geometries and for three different cavity lengths.
Consistent with the results of Fig.~\ref{fig:delF.L.D}(a), $\Delta F_{\rm c}$
increases with decreasing cavity length at any given $D$. In addition, the trend
with regard to channel shape (i.e. $\Delta F_{\rm c}$ for triangular channels is greater 
than those for square channels, which in turn is greater than those for cylindrical
channels) still holds. In the case of $L$=59, where the effects of longitudinal
confinement are negligible for this polymer length ($N$=201), the polymer behaves
simply as one confined to an infinitely long channel.  For this confinement,
the de~Gennes blob model predicts a confinement free energy that scales as 
$\Delta F_{\rm }\sim D^{-1/\nu} \approx D^{-1.70}$ (for $\nu\approx 0.588$).
We have fit the $L$=59 data to a power law $\Delta F_{\rm }\sim D^{-\beta}$, 
and the fitting curves overlaid on the data in Fig.~\ref{fig:delF.L.D}(b)
show that the data does indeed exhibit power-law behaviour. However, the measured 
scaling exponents of $\beta=1.82\pm 0.03$ for triangular channels, $\beta=1.87\pm 0.03$ 
for rectangular channels, and $\beta=1.87\pm 0.03$ for cylindrical channels
deviate somewhat from the predicted value. It is unlikely that this discrepancy arises
from confinement cavities that are insufficiently wide since the condition that $D\gtrsim 10$ 
for blob-model scaling behaviour to emerge noted in a previous study\cite{kim2013elasticity}
is satisfied here. Instead, it arises from the fact that the chains are insufficiently 
long, leading to a violation of the condition that the number of blobs satisfies 
$n_{\rm b}\gg 1$, which is also necessary to recover the predicted scaling.

\begin{figure}[!ht]
\begin{center}
\vspace*{0.2in}
\includegraphics[width=0.45\textwidth]{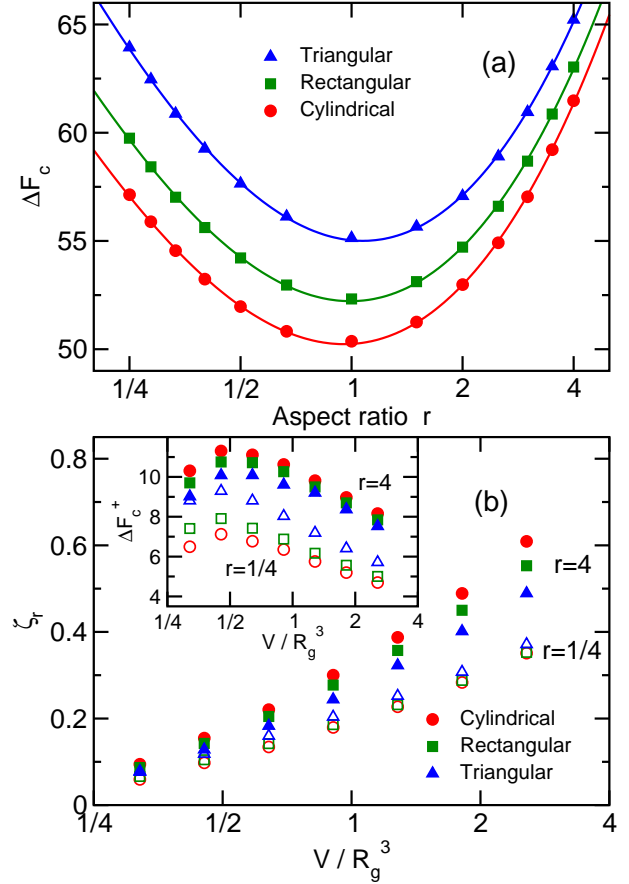}
\end{center}
\caption{(a) Confinement free energy $\Delta F_{\rm c}$ vs cavity shape anisometry, $r$.
Results are shown for polymer of length $N$=201 and for cavities of various geometry, 
each with a volume of $V$=1000 (i.e. $V/R_{\rm g}^3$=0.67). (b) Relative difference in 
the confinement free energy $\zeta_{\rm r}$ vs scaled cavity volume $V/R_{\rm g}^3$, where 
$\zeta_{\rm r}\equiv (\Delta F_{\rm c}(r)$$-$$\Delta F_{\rm }(r$=$1))/\Delta F_{\rm c}(r$=$1)$
and where $R_{\rm g}$ is the radius of gyration of a free polymer. 
Data are shown for anisometry ratios of $r$=4 and $r$=0.25 for triangular, rectangular and
cylindrical cavities and for polymers of length $N$=201. The solid symbols correspond to
$\zeta_{4}$ and the open symbols are for $\zeta_{0.25}$.
The inset shows the absolute difference in confinement free energy,
$\Delta F_{\rm c}^\dag(r)$$\equiv$$ \Delta F_{\rm c}(r)$$-$$\Delta F_{\rm }(r$=$1)$
vs $V/R_{\rm g}^3$ using the same data.
}
\label{fig:delF.ratio}
\end{figure}

Figure~\ref{fig:delF.ratio}(a) shows the variation of $\Delta F_{\rm c}$ with the
cavity aspect ratio $r\equiv L/D$ at fixed volume for three different cavity shapes.
Data are shown for a volume of $V$=1000, which corresponds to $V/R_{\rm g}^3=0.64$,
where $R_{\rm g}=11.60$ is the radius of gyration for an unconfined $N$=201 polymer.
At all aspect ratios, the same patterns noted in Fig.~\ref{fig:delF.L.D} are observed, 
i.e, the confinement free energy is greatest for triangular cavities and lowest for cylindrical 
cavities.  Notably, for all three geometries, $\Delta F_{\rm c}$ is a minimum at an aspect ratio
of $r$=1, i.e. for equal lateral and longitudinal cavity dimensions. The same trend
is observed for other cavity volumes (data not shown).  The curves
are approximately symmetric about $r$=1 on a logarithmic scale. This indicates that
the confinement free energy for aspect ratios of $r$ and $1/r$ appear to be roughly equal,
though some degree of asymmetry is evident. This asymmetry is more visible in
Fig.~\ref{fig:delF.ratio}(b), which shows the relative difference in the confinement free energy 
$\zeta_{\rm r}\equiv (\Delta F_{\rm c}(r)$$-$$\Delta F_{\rm }(r$=$1))/\Delta F_{\rm c}(r$=$1)$
as a function of cavity volume for the cases of $r$=1/4 and $r$=4. For all cavity geometries
$\zeta_4 > \zeta_{1/4}$. This is an illustration of the general trend that 
``prolate'' cavities ($L>D$) have a higher confinement free energy than ``oblate''
cavities ($L<D$). This trend is consistent with results of our previous study that examined
translocation into ellipsoidal cavities\cite{polson2015polymer} and demonstrates that it is
a generic effect independent of the details of the cavity shape.  The inset shows the absolute
difference, $\Delta F_{\rm c}^\dag(r)$$\equiv$$ \Delta F_{\rm c}(r)$$-$$\Delta F_{\rm }(r$=$1)$
for $r=4$ and $r=0.25$ and is a measure of the degree of asymmetry of the free energy for 
``prolate'' and ``oblate'' cavities of the same volume.  Interestingly, $\Delta F_{\rm c}^\dag$ 
exhibits a maximum near $V/R_{\rm g}^3\approx 0.5$.

A naive application of the approximation borrowed from the theory of semidilute polymer
solutions that the free energy is proportional to the number of blobs, i.e.
$F_{\rm c}/k_{\rm B}T=V/\xi^3$ where the correlation length scales as $\xi=\phi^{-0.77}$,
suggests that $F_{\rm c}$ should depend only on the cavity volume and not its shape.
The observed dependence of the confinement free energy on the cavity shape is most likely 
a result of the breakdown of this approximation in the limit of small cavities where $\xi$ is of 
the order of one or more cavity dimensions.  In the case of $V$=1000 and $N-1$=200 for the 
data in Fig.~\ref{fig:delF.ratio}, full insertion of the polymer leads to a volume fraction of 
$\phi_{\rm c}=0.1047$, and thus a correlation length of approximately $\xi=\phi^{-0.77}\approx 5.7$.
In the case of cylindrical cavities where $r$$\equiv$$L/D$=$1$, $D$=$L$=10.9. However, for $r$=4,
$D$=6.8 and $L$=27.3, while for $r=1/4$, $D$=17.2 and $L$=4.3. In these latter cases,
the smallest dimension $l_{\rm min}\equiv {\rm min}(L,D)$ is very close to the estimated blob size. 
Evidently, in the regime where $l_{\rm min}={\cal O}(\xi)$,
an increase in the ratio $\xi/l_{\rm min}$ leads to an increase in the free energy.
As the volume decreases and the volume fraction increases, the blob size decreases. Thus,
the effect is expected to be less significant, consistent with the observed decrease in 
$\xi_{\rm r}$ with decreasing $V$ in Fig.~\ref{fig:delF.ratio}.

\subsection{Translocation into tapered confinement spaces}
\label{subsec:cone}

Let us now consider the case of translocation into spaces with tapered geometries, 
illustrated in Fig.~\ref{fig:illust}(b). We consider first the case of a cavity 
shaped as a truncated pyramid. As noted earlier, this choice is relevant to previous 
experimental studies that have employed a pore-cavity-pore device to study DNA translocation 
into and out of pyramidal cavities.\cite{pedone2011pore,langecker2011electrophoretic} 
Here we examine the effects of size, shape and nanopore location (i.e. at the apex
or the base) on the translocation free energy. In our calculations, we fix
truncation section width to $D_{\rm a}=2$.  

\begin{figure}[!ht]
\begin{center}
\vspace*{0.2in}
\includegraphics[width=0.45\textwidth]{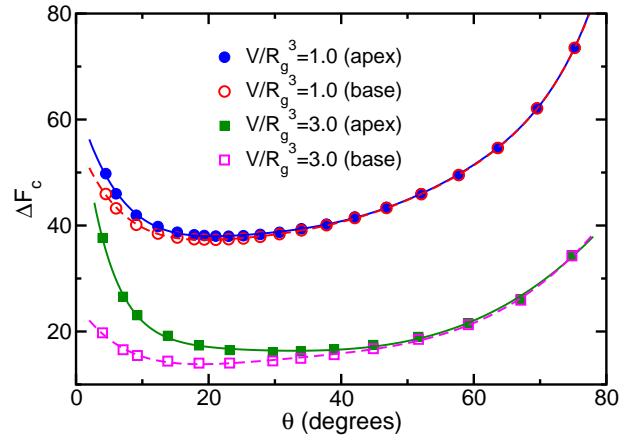}
\end{center}
\caption{Confinement free energy $\Delta F_{\rm c}$ vs taper angle $\theta$ for polymer 
translocation into a truncated pyramid. The taper angle of the pyramid is labeled in 
Fig.~\ref{fig:illust}(b).  Results are shown for a $N$=201 polymer for various cavity volumes, each 
for translocation through a pore in the pyramid base and through a pore at the pyramid apex.
The solid lines are guides for the eye.}
\label{fig:pyr.angle}
\end{figure}

Figure~\ref{fig:pyr.angle} shows the 
variation in confinement free energy $\Delta F_{\rm c}$ with the taper angle $\theta$,
which is illustrated in Fig.~\ref{fig:illust}(b). Results are shown for a polymer
of length $N$=201 entering the cavity from either the base or the apex of the
pyramid. We also consider two different cavity volumes. For all values of the
taper angle, the confinement free energy increases with decreasing volume, as expected.
In each case there is a broad minimum of $\Delta F_{\rm c}$ with respect to $\theta$.
This qualitatively similar to the trend observed in Fig.~\ref{fig:delF.ratio}(a) for
the aspect ratio in the case of constant cross-sectional area geometries. This is
not surprising, as the value of the taper angle $\theta$ determines the base/height
ratio of the pyramid, which is the equivalent of the aspect ratio for this type
of geometry. As a reference, the base/height ratio is unity when 
$\theta=\tan^{-1}(0.5)=26.6^\circ$. The location of the minimum is 
$\theta_{\rm min}\approx 20^\circ$ for all cases except for the case of $V/R_{\rm g}^3=3$
and translocation through the apex. 

A more notable feature is the contrast between apex- and base-entry translocation. 
At high density, $\Delta F_{\rm c}$ for the two cases converge at large $\theta$ 
(i.e ``squat'' pyramids). However, as $\theta$ decreases and the pyramids become ``taller'',
$\Delta F_{\rm c}$ for apex-entry translocation becomes increasingly greater than that for 
base-entry.  This trend is connected to the phenomenon of entropic depletion in the corners 
of the pyramid.  Recall that this depletion effect was also the cause of the difference in 
the values of $\Delta F_{\rm c}$ for cavities of different cross-section shapes shown in 
Figs.~\ref{fig:delF.L.D} and \ref{fig:delF.ratio}. We propose the following explanation for the 
observed trends.  As $\theta$ decreases, the apex becomes sharper and the polymer that enters 
through the base avoids occupying the region near the apex. However, apex-entry translocation 
necessarily constrains a portion of the polymer to remain in the apex region in opposition to the 
tendency for depletion. The strong confinement of that part of the polymer ultimately leads 
to a reduction in conformational entropy and thus a higher free energy relative to the case 
of base-entry translocation where depletion in the apex does occur. Depletion for the base-entry 
case increases as the apex narrows, and thus the difference between the free energies grows 
with decreasing $\theta$. For large $\theta$ (i.e. a wide apex), the entropic depletion for 
base-entry is negligible, and thus $\Delta F_{\rm c}$ for the two cases converge in this limit.
As $V$ decreases (i.e. as the density increases) the monomers are likely pushed deeper
into all the corners of the pyramid.  Thus, entropic depletion near the apex in the case of 
base-entry translocation is reduced and the difference between $\Delta F_{\rm c}$ for the two 
cases lessens even for small $\theta$, as is evident by a comparison of the results for the 
two volumes in Fig.~\ref{fig:pyr.angle}. A rigorous test of this explanation would benefit
from future measurement of the density distribution in the cavity upon variation in its size 
and shape.

\begin{figure}[!ht]
\begin{center}
\vspace*{0.2in}
\includegraphics[width=0.45\textwidth]{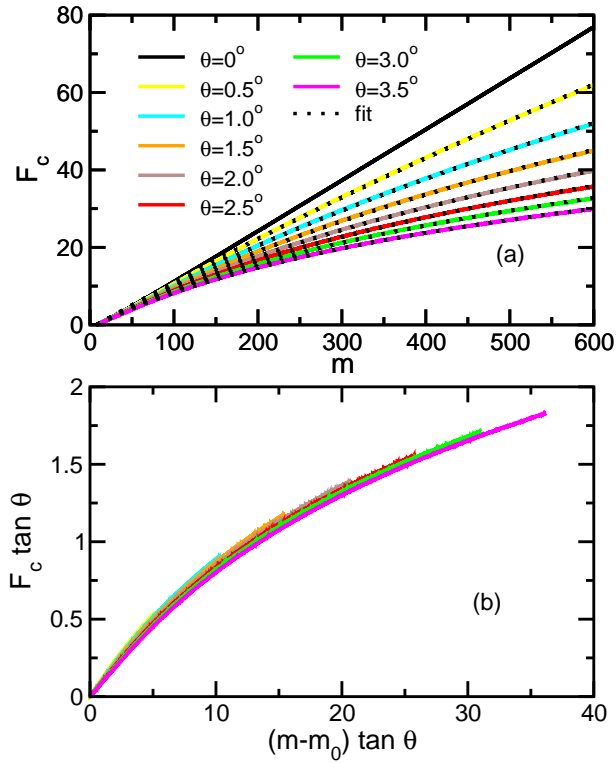}
\end{center}
\caption{(a) Confinement free energy $F_{\rm c}$ vs degree of translocation $m$ for 
translocation into an infinitely long truncated cone. Results are shown for $N=601$
for different values of the cone angle $\theta$. The diameter of the truncation 
cross section containing the pore is $D_0$=9. Overlaid on the solid curves for 
simulation data are fits using Eq.~(\ref{eq:Fcone2}).  (b) The same data as in 
(a) plotted on scaled and shifted axes. The translocation coordinate shift value 
is $m_0=8$, as explained in the text.}
\label{fig:cone}
\end{figure}

Now we consider the case of translocation into a very narrow and gradually tapered 
space. We choose $L$=$\infty$ and a circular cavity cross section, i.e. an infinitely
long cone. The free energy and dynamics of polymers in such conical spaces have been
the subject of other theoretical and simulation studies,
\cite{su2011entropically,nikoofard2013directed,nikoofard2015flexible,kumar2018polymer} 
and the behaviour of DNA in conical channels has been examined in experimental 
studies.\cite{peters2010confining} Figure~\ref{fig:cone} shows the confinement free energy 
$\Delta F_{\rm c}$ of a flexible polymer translocating into an infinitely long cone. The free 
energy is calculated as the difference $F_{\rm c}(m)\equiv F_{\rm cone}(m)-F_{\rm 0}(m)$ 
between the free-energy function for the translocation into the cone, $F_{\rm cone}(m)$, 
and that for the planar geometry, $F_0(m)$. The nanopore is located on a truncation surface 
cross-section of diameter $D_0=6$.

To analyze the results, we follow the approach taken in Ref.~\onlinecite{nikoofard2015flexible}
and derive an expression for $F_{\rm c}(m)$ using the de~Gennes blob model. First note that 
the diameter of the cone is given by $D(z)=D_0 + 2z\tan\theta$, where $\theta$ is the 
taper angle, illustrated in Fig.~\ref{fig:illust}, and $z$ is the distance from the pore 
along the central axis of the cone.  The blob size of the portion of the polymer confined 
in the cone is $\xi(z)\approx D(z)$, and the number of monomers in each blob is 
$g(z)\sim (\xi(z))^{1/\nu}$. The number of monomers $dn$ in a slice of thickness $dz$ is 
$dn(z)\sim (g(z)/\xi(z))dz$, and so the linear density of monomers along the cone, 
$\lambda(z)\equiv dn/dz$ scales as $\lambda(z)\sim (\xi(z))^{1/\nu-1}$. If the extension 
of the $m$ monomers of the polymer in the cone along $z$ is $R_{||}$, it follows that 
$m\sim \int_0^{R_{||}} \lambda(z) dz$. Thus,
\begin{eqnarray}
m = \frac{B}{\tan\theta} \left[ (D_0 + 2R_{||}\tan\theta)^{1/\nu} - D_0^{1/\nu}\right],
\label{eq:Nlambda}
\end{eqnarray}
where $B$ is a proportionality factor of order unity. In addition, the free energy due
to the tube confinement can be determined by noting that each blob contributes of the order 
of $kT$ to the confinement free energy. In the case of a continuously varying blob length, 
this implies that 
$F_{\rm c}/kT \sim \int_0^{R_{||}} [\xi(z)]^{-1}dz$. It follows that
\begin{eqnarray}
\frac{F_{\rm c}}{kT} = \frac{A}{\tan\theta} \left[ \ln (D_0+2R_{||})\tan\theta) - \ln D_0\right],
\label{eq:Fcone1}
\end{eqnarray}
where $A$ is another proportionality factor of order unity. Solving Eq.~(\ref{eq:Nlambda}) for
$R_{||}$ and substituting this expression into Eq.~(\ref{eq:Fcone1}), we find that 
\begin{eqnarray}
\frac{F_{\rm c}(m)}{kT} & = & \left(\frac{A}{\tan\theta}\right)\ln\left[
\frac{2}{D_0}\left(\frac{(m-m_0)}{B}\tan\theta + D_0^{1/\nu}\right)^\nu  -1 \right]
\nonumber \\
\label{eq:Fcone2}
\end{eqnarray}
Note in the final step we have made the substitution $m\rightarrow m-m_0$. This shift in 
the translocation is required to correct for the fact that $F_{\rm c}$ is otherwise 
predicted to increase monotonically for $m\geq 0$. In practice, this is not the case, 
as several monomers must first enter the cone from the pore before the effects of lateral 
confinement are felt. This value is expected to increase with $D_0$.  For $D_0=9$, we 
find that $F_{\rm c}$ is zero until $m$ has a threshold value of $m_0=8$.  
Equation~(\ref{eq:Fcone2}) predicts that the free energies for all cone angles should 
fall on a universal curve when plotting $F_{\rm c}\tan\theta$ vs $(m-m_0)\tan\theta$. 
Figure~\ref{fig:cone}(b) shows that the data come close to collapse on such a universal curve, 
though small discrepancies remain. Overlaid on the calculated curves in 
Fig.~\ref{fig:cone}(a) are fits using Eq.~(\ref{eq:Fcone2}). Generally, the quality of each 
fit is excellent.  However, we note that the fitting parameter values for the fits for each 
angle vary in the range $A=1.90-2.14$ and $B=0.487-0.502$. This small variation is consistent 
with the good but imperfect data collapse for the scaled data in the Fig.~\ref{fig:cone}(b).

\subsection{Effects of polymer stiffness}
\label{subsec:stiffness}

We now examine the effects of polymer stiffness on the translocation free energy. Given 
the rich scaling behaviour expected for the confinement free energy of semiflexible
polymers upon variation in the persistence and contour lengths as well as the cavity
dimensions,\cite{sakaue2018compressing} we choose to focus on two important limiting 
cases and defer a more complete exploration of parameter space to a future study. In 
particular, we choose to examine translocation cavities with: (1) an aspect ratio of 
$r\equiv L/D=1$, and (2) and aspect ratio of $r\gg 1$.

\begin{figure}[!ht]
\begin{center}
\vspace*{0.2in}
\includegraphics[width=0.45\textwidth]{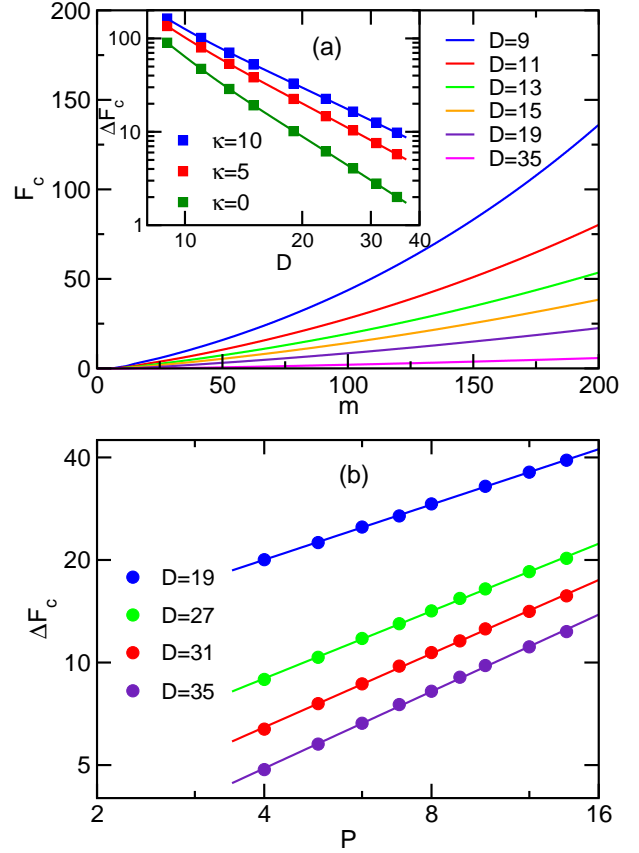}
\end{center}
\caption{(a) Confinement free energy, $F_{\rm c}$ vs $m$ for a semiflexible polymer of 
contour length $N$=201 and stiffness $\kappa$=5 (i.e. persistence length $P$=5) translocating 
into a cylindrical cavity with an aspect ratio of $D/L$=1. Results are shown for various values 
of cavity sizes.  The inset shows the 
$\Delta F_{\rm c}$ vs $D$ for polymers of stiffness $\kappa$=10, 5 and 0 (i.e. full flexible). 
Fits to $\Delta F_{\rm c}\sim D^{-\gamma}$ in the region $D\geq 14$ yields exponents
$\gamma$=2.08, 2.35 and 2.80 for $\kappa$=10, 5 and 0, respectively.
(b) $\Delta F_{\rm c}$ vs persistence length $P$ for $N$=201 and $D/L$=1. Results are shown
for four different cavity sizes. The solid lines are fits to $\Delta F_{\rm c}\sim P^\beta$,
where $\beta$=0.54, 0.66, 0.72 and 0.75 for $D$=20, 28, 32 and 36, respectively.  }
\label{fig:delF.D.kappa}
\end{figure}

Figure~\ref{fig:delF.D.kappa}(a) shows the confinement free-energy function $F_{\rm c}(m)$ 
for a $r$=1 cavity in the case of a $N$=201 polymer of persistence length $P$=5. 
Results for various cavity sizes are shown. The key trend is an increase in the
confinement free energy as $D$ decreases. There are two contributions to this effect.
First, as in the case of flexible polymers, a reduction in the cavity size will reduce the
number of available conformations of the polymer, thus reducing the entropy. Second, as
cavity size is decreased the polymer is increasingly forced to bend, leading to an
increase in the mean bending energy and thus the free energy. The inset of 
Fig.~\ref{fig:delF.D.kappa}(a) shows the variation of $\Delta F_{\rm c}$ with $D$ for 
semiflexible polymers of stiffness $\kappa$=5 and 10, as well as for a fully flexible polymer 
($\kappa$=0). In the range of $D=9-35$, the data appear to scale as
$\Delta F_{\rm c}\sim D^{-\gamma}$, though there is a slight deviation from power-law
scaling for small cavities with $D\lesssim 12$. Fits in the domain $D \geq 13$ yield 
exponents of $\gamma=2.04\pm 0.02$ for $\kappa$=10, $\gamma=2.30\pm 0.03$ for $\kappa$=5, 
and $\gamma=2.76\pm 0.03$ for a fully flexible polymer. Figure~\ref{fig:delF.D.kappa}(b) 
shows the variation of $\Delta F_{\rm c}$ with $P$ in the domain $P=4-15$ for cavities of 
various size. Again, the scaling of data appears to follow a power law of the form 
$\Delta F_{\rm c}\sim P^\beta$, where fits to the data yield scaling exponents of 
$\beta$=0.54, 0.66, 0.72 and 0.75 for $D$=19, 27, 31 and 35, respectively. 

If the system was in a well-defined scaling regime such that the confinement free energy satisfied 
$\Delta F_{\rm c}\sim D^{-\gamma} P^\beta$, the value of $\gamma$ would not depend on $P$, nor
would the value of $\beta$ change with $P$. As noted above, however, such dependencies are
observed.  One possibility is that the fits have merely yielded effective exponents in a 
cross-over region between well-defined scaling regimes. To clarify this issue let us 
consider the theoretical studies of Sakaue,\cite{sakaue2007semiflexible,sakaue2018compressing} 
who has predicted a number of free energy scaling regimes for semiflexible polymers in closed spaces 
upon variation in the polymer contour and persistence lengths and the cavity dimensions. 
The boundaries between the regimes depend on the polymer length, $N=L_{\rm c}/\sigma$, 
cavity size $D$, and the ratio $p\equiv l_{\rm K}/\sigma=2P/\sigma$, where $l_{\rm K}$ is 
the Kuhn length. (We use the notation of the present article and the convention for the
definition of $N$ in Ref.~\onlinecite{sakaue2007semiflexible} rather than 
Ref.~\onlinecite{sakaue2018compressing}.) In our calculations, $N\ll p^3$ in all cases and
thus the scaling of the fluctuating semidilute regime (regime~I from 
Ref.~\onlinecite{sakaue2007semiflexible} and $\rm F^3_{~0}$ in 
Ref.~\onlinecite{sakaue2018compressing}) is not expected to relevant to our results.
(It does, however, provide the correct scaling with respect to $D$ for flexible polymers.)
On the other hand, we note that $N\approx p^2$ and $D\gtrsim p$. As a consequence, the system 
is expected to be in a region of parameter space near the convergence of the following four scaling
regimes illustrated in Fig.~2 of Ref.~\onlinecite{sakaue2007semiflexible}: (a) the mean-field 
semidilute regime (regime~II), where $\Delta F_{\rm c}/k_{\rm B}T \sim N^2 D^{-3} P^0$; 
(b) the liquid crystalline regime (regime~III), where 
$\Delta F_{\rm c}/k_{\rm B}T \sim N^1 D^{0} P^{-1}$;
(c) the ideal chain regime (regime~IV), where $\Delta F_{\rm c}/k_{\rm B}T \sim N^1 D^{-2} P^{1}$;
(d) the bending regime (regime~V), where $\Delta F_{\rm c}/k_{\rm B}T \sim N^1 D^{-2} P^{1}$.
(Note that in Ref.~\onlinecite{sakaue2018compressing} these regimes are labeled $\rm M^3_{~0}$,
$\rm N^3_{~0}$, $\rm I_0$, and $\rm S_0$, respectively.) The effective scaling exponents
extracted from fits to our data are qualitatively inconsistent with regime~III and so
suggest the system is in a transition region between regime~II and regimes IV/V (note that
the latter two satisfy the same scaling). In particular, the measured $\gamma$ lies between
the values for those regimes of $\gamma=2$ and $\gamma=3$.  Likewise, the measured value 
of $\beta$ lies between the values for those regimes of $\beta=0$ and $\beta=1$.
Note as well that increasing $D$ leads to a measured scaling exponent $\beta$ closer to the
value of $\beta=1$ predicted for regime~IV, consistent with the trends of the 
scaling regime boundaries in Fig.~2 of Ref.~\onlinecite{sakaue2007semiflexible}.
Likewise, increasing $\kappa$, and therefore $P$, leads to an exponent $\gamma$ that 
tends toward the value of $\gamma=2$ for regime IV, which is also qualitatively consistent 
with the trends for the confinement regimes predicted in Ref.~\onlinecite{sakaue2007semiflexible}.

One feature of the present system that complicates a comparison
of our results with Sakaue's predictions is the fact that here $\Delta F_{\rm c}$ 
represents the confinement free energy of a polymer that is effectively tethered to 
one wall of the cavity (because one monomer still lies in the pore when $m=N-1$). 
As noted in Ref.~\onlinecite{polson2019free} for the case of {\it flexible chains} 
the free-energy cost of such tethering is only 1--2 $k_{\rm B}T$ for polymers of
comparable length and packing fraction as that considered here.
Thus, the effect on on the free energy is expected to be negligible.
However, for {\it stiff} chains this tethering also leads to an orientational anchoring to the
cavity wall with the chain contour tending to be perpendicular to the wall at the 
effective tethering point (i.e. the pore). It is possible that this feature will alter the confinement
free energy in a manner that could further perturb the scaling properties of the free energy.

Obviously, this analysis of our simulation results represents nothing like a rigorous test of
the predictions of Refs.~\onlinecite{sakaue2007semiflexible} and \onlinecite{sakaue2018compressing}.
At a minimum, such a test requires using polymers of substantially greater length in order 
for a system to lie unambiguously in a well-defined scaling regime rather than in a transition 
region.  Nevertheless, our results do at least provide some tentative and indirect supporting 
evidence for the scaling predictions. 

\begin{figure}[!ht]
\begin{center}
\vspace*{0.2in}
\includegraphics[width=0.45\textwidth]{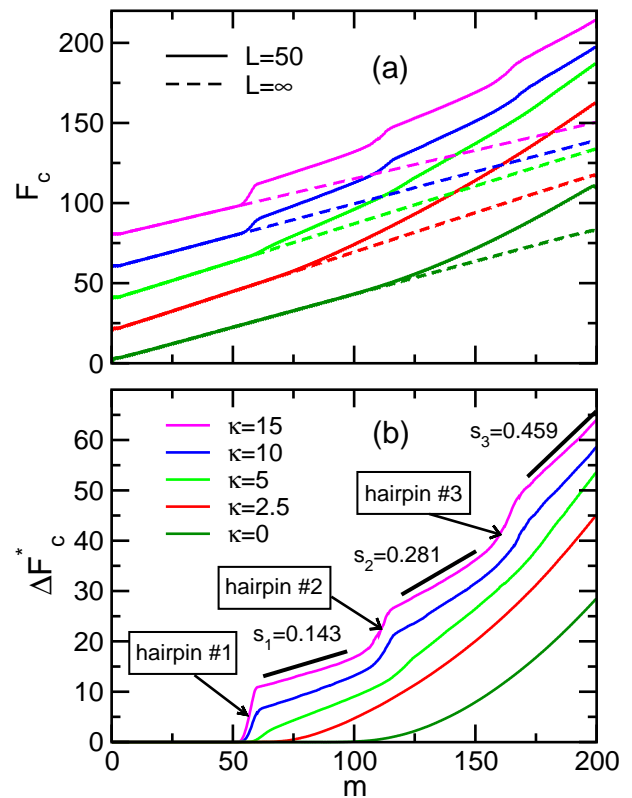}
\end{center}
\caption{(a) Free energy for translocation of a semiflexible polymer of length $N$=201 into
a cylindrical channel of width $D=4$.  Results are shown for various degrees of polymer
stiffness for cylinders of length $L$=25 and $L$=$\infty$. The curves are vertically shifted
and spread out along the vertical axis to minimize overlap for clarity. (b) Difference in the 
free energy, $\Delta F_{\rm c}^*\equiv F_{\rm c}(m;L=50)-F_{\rm c}(m;L=\infty)$ vs $m$.  
The black line segments in the inset are linear fits (shifted upward) to the linear sections 
of the $\kappa=15$ curve. The slopes of the fits are each labeled, as are the locations 
of each hairpin that separate the linear regions. }
\label{fig:delF.N201.R2.5.L50-500}
\end{figure}

Having first considered the case of translocation of semiflexible polymers into cavities with 
aspect ratios of $r$=1, let us now consider the case of highly asymmetric cavities with $r\gg 1$. 
In addition, we also focus on very narrow cavities such that $D\lesssim P$. In the case of 
very long cavities where $L>L_{\rm c}$ (where $L_{\rm c}$ is the polymer contour length),
this corresponds to the Odijk or backfolded Odijk regimes. However, for $L<L_{\rm c}$
the finite cavity length is expected to produce different conformational behaviour.
Figure~\ref{fig:delF.N201.R2.5.L50-500} (a) shows free-energy functions for a semiflexible 
polymer of length $N$=201 entering a cylindrical cavity of width $D$=4. Results are shown for 
polymers of varying degrees of rigidity ranging from fully flexible to $\kappa$=15. In each case,
curves are shown for a cavity of length $L$=50 (solid curves) and $L$=$\infty$ (dashed
curves).  Functions for different $\kappa$ are vertically shifted relative to each other
for clarity. The curves for each $\kappa$ initially overlap at low $m$, where
the translocated part of the polymer is not sufficiently long to feel the effects
of longitudinal confinement for $L$=50. As $m$ increases further, the free-energy functions
for the longitudinally confined systems diverge from the $L$=$\infty$ curves at the point
where the polymer makes contact with the confining cap. The free energy increase with
respect to the $L$=$\infty$ case arises from the reduction in the conformational entropy 
resulting from this additional confinement. As expected, the value of $m$ at which the
divergence occurs decreases with increasing polymer stiffness. This follows from the 
fact that stiffer polymers are more elongated in the tube, and thus fewer translocated
monomers are required before the polymer reaches the cap. For $P<D$, $F$ increases
smoothly with positive curvature. However, for $P\gtrsim D$, curves are qualitatively 
different in that $F$ increases markedly in steps with linear regimes in between.
The size of the step increases with $\kappa$ and the locations of the steps appear to
converge to values of $m$ that are integer multiples of approximately $\Delta m$=55. 

\begin{figure}[!ht]
\begin{center}
\vspace*{0.2in}
\includegraphics[width=0.40\textwidth]{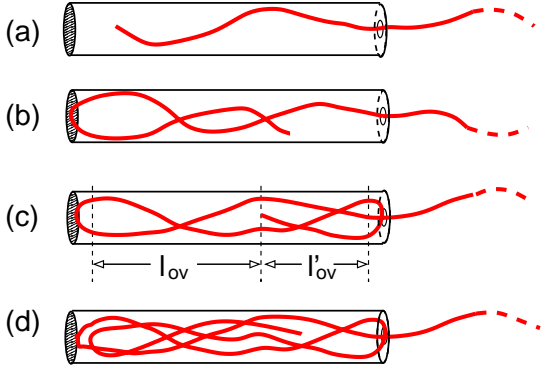}
\end{center}
\caption{Illustration of the backfolding domains present for the data of $\kappa$=15
in Fig.~\ref{fig:delF.N201.R2.5.L50-500}. Parts (a), (b), (c) and (d) show domains
with 0, 1, 2 and 3 backfolds, respectively.}
\label{fig:illust2}
\end{figure}

The origin of the step behaviour is straightforward. In the case of the more flexible chains, 
increasing $m$ will typically cause a gradual increase in the density of monomers, which will 
likely maintain their linear organization along the channel, i.e. a longitudinally uniform 
compression. At high density (i.e. high $m$), the translocated portion of the polymer may 
lose its linear organization as it forms backfolds, but no obvious signature in $F_{\rm c}(m)$ is 
expected. By contrast, a stiffer polymer is expected to initially undergo uniform longitudinal 
compression until the point that it becomes more favourable to pay the cost of forming a 
hairpin turn to minimize $F_{\rm c}$. The steps in $F_{\rm c}(m)$ are signatures of these backfolds, 
and the 
increase in $F_{\rm c}$ in each step corresponds to the free energy of hairpin formation. As expected, 
the hairpin free energy increases with increasing polymer stiffness, principally as a result 
of the greater energy requirement to form the hairpin, though it should be noted that there 
can be a significant entropic contribution to the hairpin free energy as 
well.\cite{polson2017free} In the limit of large $\kappa$, the translocated section of
the polymer becomes highly aligned with the channel and the end of the polymer is expected
to make contact with the confining cap when the contour length of this section is close to
the length of the cylinder, i.e. when $m\approx 50$ for a tube of length $L$=50. Small
lateral fluctuations in the polymer conformation mean that a slightly longer translocated
polymer length is required before the polymer end reaches the cap. In the case of $\kappa=15$
the first step occurs at $m\approx 55$.  The successive steps each correspond to increasing
numbers of hairpin backfolds. For example, the step at $m\approx 110$ corresponds to the formation
of a second backfold located on the wall where the nanopore is located, and so on.
The process of formation of successive backfolds as translocation proceeds is illustrated
in Fig.~\ref{fig:illust2}.

As noted above, for sufficiently high $\kappa$ the regions between the steps exhibit a 
linear variation of $F_{\rm c}$ with $m$. Furthermore, the slope of these linear regions increase
as the number of backfolds increases. To illustrate this, we plot the difference 
$\Delta F^*_{\rm c} \equiv F_{\rm c}(m;L=50)-F_{\rm c}(m;L=\infty)$ in 
Fig.~\ref{fig:delF.N201.R2.5.L50-500}(b).  The increase in the slopes for $\kappa=15$ is 
more clearly evident. Linear fits to each of these regions are shown as black line segments 
in the figure (shifted vertically for clarity). The slope of each fit, labeled in the figure, 
increases as the number of hairpin backfolds increases. 

To explain the origin of the linear regions and the variation of the slope with the number 
of hairpins, we use a theoretical approach that we previously employed to explain backfolding 
of semiflexible polymers in infinitely long channels.\cite{polson2017free,polson2018free} 
This model relies on the Odijk regime condition that $P\gg D$, which is marginally satisfied
for $\kappa$=15 and $D$=4. In the case of $L$=$\infty$ there is no backfolding and thus
no overlap of the polymer along the tube. In this case the confinement free energy is
proportional to the number of deflection segments, each of which contributes of order $kT$.
Since the number of deflection segments is proportional to the number of translocated
monomers, it follows that $F_{\rm c}\propto m$. In a backfolded regime where overlap is present,
there are additional contributions to the free energy from the hairpin and excluded volume 
interactions between the deflection segments. After a complete hairpin is formed,  
increasing $m$ simply increases the degree of overlap. Modeling the deflection segments
as hard rods of length $l_{\rm d}\sim D^{2/3}P^{1/3}$, the interaction free energy
of $N$ such rods is $F_{\rm int}\approx l_{\rm d}^2 w N^2\langle|\sin\gamma|\rangle/V$,
where $V$ is the volume occupied by the segments and where the angle $\gamma$ between
a pair rods satisfies $\langle|\sin\gamma|\rangle \approx D/l_{\rm d}$ in the case where
they are highly aligned in the cylinder. The overlap volume can be written $V\sim l_{\rm ov}D^2$,
where $l_{\rm ov}$ is the length of the overlap regime along the tube. Thus, the
interaction free energy is $F_{\rm int} \sim l_{\rm d}wN^2/(l_{\rm ov}D^2)$. Now, in a
region where $F(m)$ is linear, there are two different overlap regions along the tube,
one with $n$ overlapping strands and one with $n^\prime=n+1$ such strands, where $n$ is 
the number of hairpins present. The number of deflection segments in each region is
$N=nl_{\rm ov}/l_{\rm d}$ and $N^{\prime}=(n+1)l^\prime_{\rm ov}/l_{\rm d}$, where $l_{\rm ov}$
and $l_{\rm ov}^\prime$ are the lengths of the two overlapping regions. (These lengths are
illustrated in Fig.~\ref{fig:illust2}(c) for the case of $n$=2.)  The total free energy
arising from interacting segments from the two regions is
\begin{eqnarray}
F_{\rm int} \sim \frac{l_{\rm d}wN^2}{l_{\rm ov}D} + \frac{l_{\rm d}w(N^\prime)^2}{l^\prime_{\rm ov}D}
\label{eq:Fint1}
\end{eqnarray}
The overlap lengths are simply related by $l_{\rm ov} = L -2h_z - l^\prime_{\rm ov}$,
where $h_z$ is the size of the hairpin along the channel. In addition, for a highly aligned
polymer, $l^\prime_{\rm ov} \sim m$. Finally, noting that the Odijk deflection length scales 
as $l_{\rm d}\sim D^{2/3}P^{1/3}$, it is straightforward to show that Eq.~(\ref{eq:Fint1})
reduces to $F_{\rm int} \sim D^{-5/3}P^{-1/3}(2n+1) m$ plus terms independent of $m$. 
Thus, the contribution to the free-energy gradient from the excluded volume interactions 
is expected to scale as 
\begin{eqnarray}
f_{\rm int}\equiv \frac{dF_{\rm int}}{dm} \sim D^{-5/3}P^{-1/3}(2n+1)
\label{eq:fint}
\end{eqnarray}
Thus, $f_{\rm int}$ is predicted to increase with the number of hairpins, consistent
with the observation in Fig.~\ref{fig:delF.N201.R2.5.L50-500}(b). For a quantitative
comparison, we use Eq.~(\ref{eq:fint}) to estimate ratios of the gradients. We find that 
$f_{\rm int}(n=2)/f_{\rm int}(n=1) \approx 1.67$ and 
$f_{\rm int}(n=3)/f_{\rm int}(n=1)\approx 2.33$. By comparison, we find the corresponding
ratios of the gradients in Fig.~\ref{fig:delF.N201.R2.5.L50-500}(b) are $0.281/0.143=1.96$
and $0.459/0.143=3.21$, respectively. Thus, the theoretical model underestimates the 
ratios of the gradients, and the discrepancy appears to grow as the number of hairpins
increases. A similar discrepancy was observed using in Ref.~\onlinecite{polson2017free}
for the ratios of overlap free-energy gradients of backfolded semiflexible polymers
confined to long cylinders in the case where a single hairpin is present and the
case of an S-loop with two hairpins. Likely origins of this discrepancy include
not sufficiently satisfying the Odijk condition $P\gg D$, treating interactions at
the second-virial level in a regime where the strands are tightly packed in a very
narrow tube, and the neglect of correlations in position and orientation of deflection
segments connected to the same hairpin.

Finally, it should be noted that the conformations characterized by multiple hairpins
separated by elongated strands of Odijk deflection segments were observed here for the case 
where the persistence length $P$ is significantly less than the contour length $L_{\rm c}$ 
of the polymer, in addition to satisfying $P>D$. Recently, qualitatively different
behaviour was observed in the case much stiffer polymers.\cite{hayase2017compressive}
In the regime where $P\approx L_{\rm c}$, compression of a polymer in a finite-length
channel resulted in the formation helical structures prior to the formation of hairpins.
In the future, it would be of interest to examine the confinement free energy in
this regime.

\subsection{Effects of Crowding Agents}
\label{subsec:crowding}

\begin{figure}[H]
\begin{center}
\vspace*{0.2in}
\includegraphics[width=0.45\textwidth]{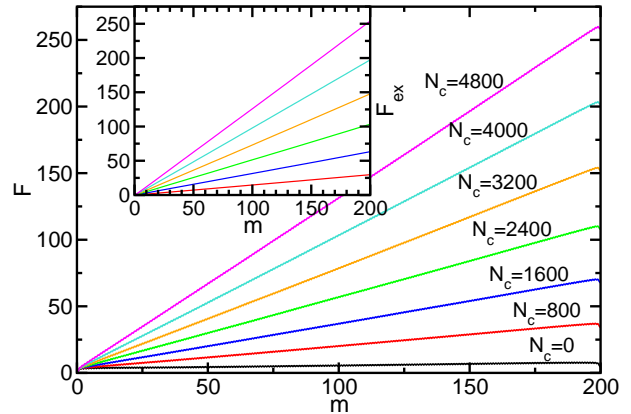}
\end{center}
\caption{Free energy for translocation of a polymer of length $N$=201 
into a cylindrical cavity of dimensions $D$=$L$=28 that is partially occupied
by monomer-sized crowding agents. Results are shown for various values of the number of
crowders, $N_{\rm c}$. The inset shows the corresponding excess free energy $F_{\rm ex}$ 
obtained from the translocation free-energy functions, where 
$F_{\rm ex}(m;N_{\rm c})\equiv F(m;N_{\rm c})-F(m;N_{\rm c}=0)$.
}
\label{fig:delF.N201.R14.L28}
\end{figure}

We now investigate the effects of crowding agents on the free-energy functions for 
polymer translocation into confined cavities. For this purpose, we choose to employ
symmetric cylindrical cavities (i.e. $D$=$L$) and consider first the case of 
monomer-sized crowding agents, i.e. $\sigma_{\rm c}$=$\sigma$=1. 
Figure ~\ref{fig:delF.N201.R14.L28} shows free-energy functions for translocation
of a $N$=201 polymer into a cylindrical cavity of dimensions $D$$=$$L$=28 whose volume
is partially occupied with $N_{\rm c}$ crowding agents, where $N_{\rm c}$ ranges
from 0 to 4800, (i.e. crowder packing fractions up to $\phi_{\rm c}$=0.146).
As expected, the free-energy cost for polymer insertion increases with increasing
crowder density. The inset of the figure shows the excess free energy, which we
define as $F_{\rm ex}(m)\equiv F(m;N_{\rm c})-F(m;N_{\rm c}=0)$; that is, 
$F_{\rm ex}(m)$ measures the variation of the free energy of the cavity/crowder
system in excess of the variation in the free energy due to confinement alone,
$F(m;N_{\rm c}=0)$. As is evident in the figure, $F_{\rm ex}$ varies linearly
with $m$. Note that cavity size is such that $D/R_{\rm g}$=2.33, where $R_{\rm g}$
is the radius of gyration of a free polymer of the same size. For smaller values
of $D/R_{\rm g}$, the same general trends are observed, though $F_{\rm ex}$ 
becomes increasingly less linear (data not shown).

\begin{figure}[!ht]
\begin{center}
\vspace*{0.2in}
\includegraphics[width=0.45\textwidth]{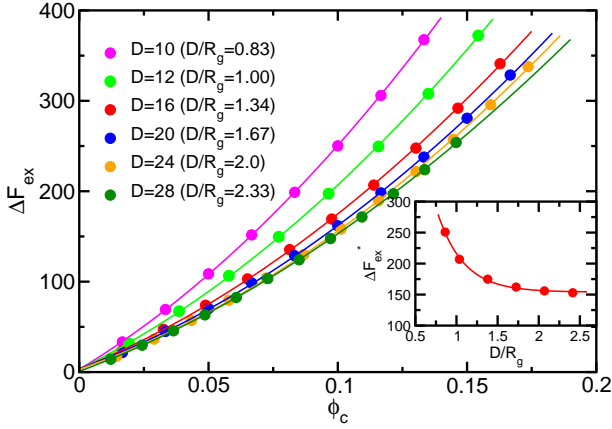}
\end{center}
\caption{ Excess free energy $\Delta F_{\rm ex}\equiv F_{\rm ex}(m=N)$ of inserting a polymer 
into a cavity partially occupied with crowding agents vs crowding agent packing fraction 
$\phi_{\rm c}$. We employ monomer-sized crowders, i.e. $\sigma_{\rm c}$=1, a 
polymer of length $N$=201, and a cylindrical cavity with $D$=$L$. Results are shown for 
several different cavity sizes, each labeled in the figure in relation to the radius of gyration 
$R_{\rm g}$ for a free polymer.  The inset shows the variation of $\Delta F_{\rm ext}^*$ with 
$D$, where $\Delta F_{\rm ext}^*\equiv\Delta F_{\rm ex}(\phi_{\rm c}=0.1)$.
}
\label{fig:delF.phic.N201}
\end{figure}

Figure~\ref{fig:delF.phic.N201} shows the variation of the excess insertion
free energy for complete polymer insertion, $\Delta F_{\rm ex}\equiv F_{\rm ex}(m=N)$,
vs crowding packing fraction $\phi_{\rm c}$ for $N$=201 and isotropic cavities with $D$=$L$.
Results are shown for different cavity sizes with $D$ ranging from 10--28
(and thus $D/R_{\rm g}$=0.83--2.33). For each cavity size, $\Delta F_{\rm ex}$ increases
monotonically with increasing crowder packing fraction. At any given packing fraction,
the excess free energy decreases monotonically with increasing cavity size and appears
to converge to a single curve for sufficiently large $D$. This is illustrated in the inset
of the figure, which shows $\Delta F_{\rm ex}$ vs $D$ for a packing fraction of 
$\phi_{\rm c}$=0.1.

\begin{figure}[!ht]
\begin{center}
\vspace*{0.2in}
\includegraphics[width=0.45\textwidth]{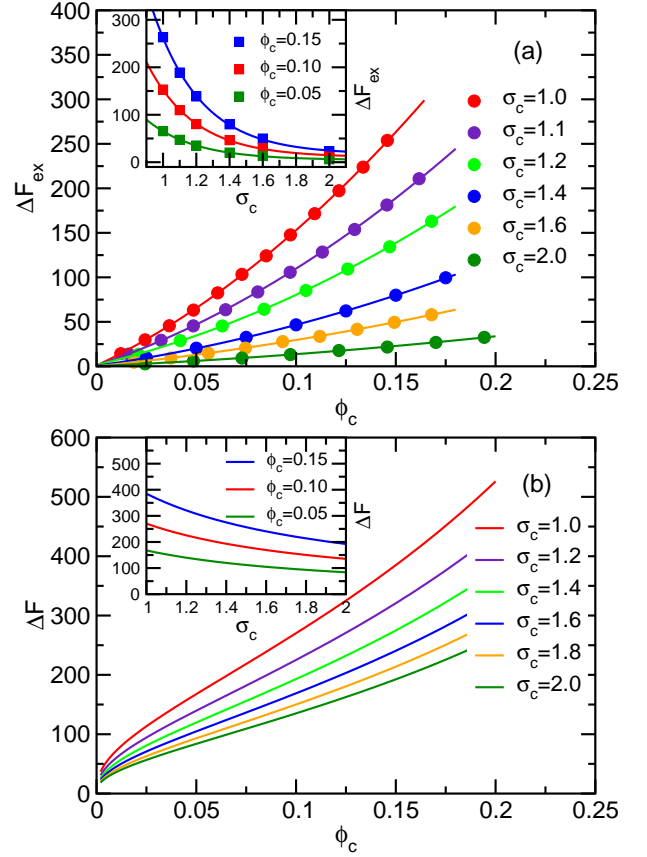}
\end{center}
\caption{(a) Excess free energy $\Delta F_{\rm ex}\equiv F_{\rm ex}(m=N)$ of inserting a polymer 
into a cavity partially occupied with crowding agents vs crowding agent packing fraction 
$\phi_{\rm c}$. Results are shown for several values of the crowder size diameter $\sigma_{\rm c}$ 
for a polymer of length $N$=201 and a cylindrical cavity of dimensions $D$=14 and $L$=28.  The 
inset shows $\Delta F_{\rm ex}$ vs $\sigma_{\rm c}$ for three different packing fractions using 
interpolations of the data in the main part of the figure. (b) Prediction of free-energy
difference $\Delta F=F(m=N)-F(m=0)$ vs $\phi_{\rm c}$ using Eq.~(\ref{eq:FLuo}). The
inset shows the prediction for $\Delta F$ vs $\sigma_{\rm c}$ for various packing fractions
using Eq.~(\ref{eq:FLuo}).
}
\label{fig:delF.size.phic}
\end{figure}

Figure~\ref{fig:delF.size.phic}(a) shows the variation of $\Delta F_{\rm ex}$ with $\phi_{\rm c}$ 
for $N$=201 and cavities with dimensions $D$=$L$=28. Results are shown for different crowder 
sizes in the range $\sigma_{\rm c}=$1.0--2.0.  For each $\sigma_{\rm c}$ considered, 
$\Delta F_{\rm ex}$ increases with increasing $\phi_{\rm c}$. More significantly, at fixed
crowder packing fraction, the excess free energy decreases monotonically with increasing
crowder size. The effect is quite pronounced, as is also evident in the inset which 
shows $\Delta F_{\rm ex}$ vs $\sigma_{\rm c}$ for various packing fractions. 

The results in Fig.~\ref{fig:delF.size.phic}(a) are relevant to the simulation study by Chen and 
Luo.\cite{chen2013dynamics} In that work, the rate of translocation was examined for a polymer
translocating between two spaces each occupied with crowding agents of different sizes but
with equal packing fractions.  They observed that a polymer initially configured with its 
center monomer in the nanopore tends to translocate into the space with the larger crowding
agents. Note that the range of packing fractions and crowder sizes they considered
($\phi_{\rm c}$=0.05--0.4, $\sigma_{\rm c}$=1--2.5) are comparable to that examined here.
However, since they used a 2-D system, a direct quantitative comparison with our results 
is not possible.  Nevertheless, their observation is qualitatively consistent with our
calculations, which predict a lower free energy for larger crowders at fixed $\phi_{\rm c}$ 
and thus a free-energy gradient that will drive translocation in that direction.
Chen and Luo note, however, that the probability that the polymer goes to the 
high-$\sigma_{\rm c}$ side exhibits a maximum upon increasing $\sigma_{\rm c}$, and
likewise the translocation rate exhibits a minimum. This is not consistent with trends
evident in the free-energy calculations. Chen and Luo attribute this to kinetic effects
due to ``bottlenecks'' related to the relative timescales of the conformational 
relaxation of the polymer and the diffusion of the obstacles. At low $\phi_{\rm c}$ and
large $\sigma_{\rm c}$, this leads to ``resisting force'' of appreciable magnitude that
effectively counteracts the entropic force, reducing the likelihood that the polymer
reaches the higher-$\sigma_{\rm c}$ side. As this is an out-of-equilibrium effect,
our free-energy calculations cannot account for this behaviour. However, our results
can be used to test the analytical model used in Ref.~\onlinecite{chen2013dynamics} to 
approximate the free-energy gradient. 

In their theoretical model, Chen and Luo assume that this entropic force exerted
by obstacles on either side of the nanopore scales $f\sim 1/R$, where $R$ is the mean 
spacing between the crowders. This assumption is inspired from a previous observation 
that a polymer ejected from a cylindrical nanochannel of diameter $R$ experiences a
driving force with the same scaling. However, since a channel of fixed shape differs
appreciably from the effective channels of fluctuating shape and size formed by
the spaces between crowders, the accuracy of this prediction is not obvious 
{\it a priori}. Following the approach in Ref.~\onlinecite{chen2013dynamics} and
adapting it to a 3-D system, it is easily shown that the mean spacing between 
crowders is given by $R\approx (\pi/6\phi_{\rm c})^{1/3}\sigma_{\rm c}-\sigma_{\rm c}$.
Noting that $f\propto dF/dm \propto 1/R$ and integrating with respect to $m$, it follows 
that 
\begin{eqnarray}
\Delta F \equiv F(N)-F(0) \propto \frac{N}{(\pi/6\phi_{\rm c})^{1/3}\sigma_{\rm c}-\sigma_{\rm c}}.
\label{eq:FLuo}
\end{eqnarray}
Figure~\ref{fig:delF.size.phic}(b) shows the predicted variation of $\Delta F$ with $\sigma_{\rm c}$
and (in the inset) with $\phi_{\rm c}$. Comparing the prediction with the simulation data
reveals that the theory correctly predicts the qualitative trends, i.e. $\Delta F$ increases
monotonically with increasing $\phi_{\rm c}$ for arbitrary crowder size, and it decreases
monotonically with increasing $\sigma_{\rm c}$ for arbitrary packing fraction. As expected,
however, the quantitative accuracy is very poor. It significantly overestimates the rate of 
increase in $\Delta F$ with $\phi_{\rm c}$ at low $\phi_{\rm c}$, and it significantly
underestimates the rate of decrease in $\Delta F$ with increasing crowder size. Doubtless,
the main cause of the discrepancy is the assumption that $f\sim 1/R$.

\section{Conclusions}
\label{sec:conclusions}

In this study, we have used computer simulations to measure the free energy of a polymer 
undergoing translocation through a nanopore into a confining cavity.  The scaling properties 
of the confinement free energy were examined with respect to the variation in several key 
system properties, including polymer length, cavity size and shape, polymer stiffness, and 
crowding from mobile crowding agents inside the cavity. These results complement and build 
on those of a previous study where we examined translocation into an ellipsoidal 
cavity.\cite{polson2015polymer} The scaling results were typically compared with predictions 
obtained using standard scaling theories of polymer physics. While the measured scaling 
exponents are generally comparable to the predicted values, discrepancies arising from 
finite-size effects persist even for the longest polymer length employed here ($N$=601). 
A more rigorous test of the theoretical predictions in the future will likely require 
simulations employing polymer lengths at least an order of magnitude larger than is
currently feasible. It will also be beneficial to consider other experimentally relevant
factors such as the effects of electric driving forces and adsorption to the inner surface
of the confining cavity.\cite{rasmussen2012translocation,polson2015polymer} 
Clearly, in the absence of such forces the gradient in the free energy tends to
drive the polymer {\it out} of the cavity. When present, however, they can provide
a decrease in the potential energy as the monomers move inside the cavity that offsets
or eliminates the loss in conformational entropy, thus driving the polymer inward.

Finally, it will be of interest to carry out additional simulations to measure and 
characterize the dynamics of polymer translocation into or out of confined cavities. 
For hard-sphere-chain polymers used here, either MC dynamics\cite{polson2013polymer} or
Discontinuous Molecular Dynamics\cite{opps2013role} simulations would be appropriate,
while use of Brownian or Langevin dynamics techniques requires the use of continuous-potential
models.  A comparison of the rates of translocation into or out of the cavity
with predictions from calculations employing the Fokker-Planck formalism will provide a means 
to delineate the regime in which translocation is a quasistatic process governed by 
the equilibrium free-energy function, as shown in previous 
work.\cite{polson2013polymer,polson2014evaluating}

\begin{acknowledgments}
This work was supported by the Natural Sciences and Engineering Research Council of 
Canada (NSERC).  We are grateful to the Atlantic Computational Excellence Network 
(ACEnet) for use of their computational resources.
\end{acknowledgments}


%

\end{document}